\documentstyle[12pt]{article}
\newcommand{\thspace}{\kern.08333em}
\newcommand{\sst}{\scriptstyle}
\newcommand{\sss}{\scriptscriptstyle}
\newcommand{\bd}{B_d^0}
\newcommand{\bdb}{\overline{B_d^0}}
\font\er=cmr10 scaled\magstep0
\def \beq{\begin{equation}}
\def \eeq{\end{equation}}
\def \Bbar{\overline{B}}
\def \Kbar{\overline{K}}
\def \s{\sqrt{2}}

\def \v#1#2{V_{#1#2}}
\def \vc#1#2{V^*_{#1#2}}
\def\roughly#1{\mathrel{\raise.3ex\hbox{$#1$\kern-.75em\lower1ex\hbox{$\sim$}}}}
\def\lsim{\roughly<}
\begin{document}
\rightline{UdeM-LPN-TH-94-195}
\smallskip
\rightline{TECHNION-PH-94-9}
\smallskip
\rightline{EFI-94-18}
\smallskip
\rightline{hep-ph/9404281}
\bigskip
\rightline{April, 1994}
\bigskip
\bigskip
\centerline{\bf MEASURING STRONG AND WEAK PHASES}
\centerline{\bf IN TIME-INDEPENDENT B DECAYS}
\bigskip
\centerline{\it Oscar F. Hern\'andez\footnote{e-mail:
oscarh@lps.umontreal.ca} {\rm and} David London\footnote{e-mail:
london@lps.umontreal.ca}}
\centerline{\it Laboratoire de Physique Nucl\'eaire}
\centerline{\it Universit\'e de Montr\'eal, Montr\'eal, PQ, Canada H3C 3J7}
\medskip
\centerline{and}
\medskip
\centerline{\it Michael Gronau}
\centerline{\it Department of Physics}
\centerline{\it Technion -- Israel Institute of Technology, Haifa 32000,
Israel}
\medskip
\centerline{and}
\medskip
\centerline{\it Jonathan L. Rosner}
\centerline{\it Enrico Fermi Institute and Department of Physics}
\centerline{\it University of Chicago, Chicago, IL 60637}
\bigskip
\bigskip
\centerline{\bf ABSTRACT}
\medskip
\begin{quote}
Flavor SU(3) symmetry implies certain relations among $B$-decay amplitudes
to $\pi\pi$, $\pi K$ and $K {\bar K}$ final states, when annihilation-like
diagrams are neglected. Using three triangle relations, we show how to
measure the weak CKM phases $\alpha$ and $\gamma$ using time-independent
rate measurements only. In addition, one obtains all the strong final-state
phases and the magnitudes of individual terms describing tree (spectator),
color-suppressed and penguin diagrams. Many independent measurements of
these quantities can be made with this method, which helps to eliminate
possible discrete ambiguities and to estimate the size of SU(3)-breaking
effects.
\end{quote}
\newpage

\centerline{\bf I. INTRODUCTION}
\bigskip

In the near future, the study of $B$ meson decays will be a crucial testing
ground for the Standard Model (SM) picture of CP violation based on phases
in the Cabibbo-Kobayashi-Maskawa (CKM) \cite{CKM} matrix $V_{\alpha i}$,
where $i = (d,s,b)$ and $\alpha = (u,c,t)$. Unitarity of the CKM matrix
implies the following triangle identity (the ``unitarity triangle'')
\beq
\v u d \vc u b + \v c d \vc c b + \v t d \vc t b = 0 ~.
\label{unitri}
\eeq
In the now-familiar Wolfenstein parametrization \cite{Wolfenstein}, only
$V_{ub}$ and $V_{td}$ have non-negligible phases, so that the angles in the
triangle are given by $\beta = -{\rm Arg} (\v t d)$, $\gamma = {\rm
Arg}(\vc u b)$, and $\alpha=\pi-\beta-\gamma$ \cite{CPangles}. The SM can
be tested by independently measuring the three angles $\alpha$, $\beta$ and
$\gamma$.

CP violation can occur in the $B$ system if two weak amplitudes contribute
to a particular decay. It is necessary that there be both a weak and a
strong phase difference between these two amplitudes in order to see an
asymmetry between the rates for $B\to f$ and ${\bar B}\to{\bar f}$. One
advantage of this type of CP violation is that it can occur in
``self-tagging'' modes such as those involving charged $B$ decays (e.g.\
$B^+ \to \pi^0 K^+$). The major disadvantage is that, since the strong
phases are unknown, measurements of CP violation in such systems will not
yield clean information about the weak CKM phases.

A potentially cleaner class of CP-violating asymmetries involves the decay
of neutral $B$ mesons to CP eigenstates such as $J/\psi K_S$ or $\pi^+
\pi^-$. In this case, CP violation arises from the interference between a
direct decay amplitude and one which proceeds via $B^0$-$\bar B^0$ mixing.
If one weak amplitude contributes to the direct decay, CKM phase
information can be extracted from measurements of these asymmetries
independent of strong final-state phases. Such measurements require the
ability to obtain time-dependent information and to ``tag'' the flavor of
the decaying $B$ meson, i.e.\ to know whether it was a $B^0$ or a $\bar
B^0$ at $t=0$.

One possible problem in the above program is that there may be more than
one weak amplitude -- in addition to tree-level decays, ``penguin''
diagrams may contribute significantly \cite{penguins,gropenguin}. In this
case, both types of CP violation are present, which complicates things
considerably -- it is no longer possible to obtain clean CKM phase
information in the same simple way. However, by using isospin relations
among several amplitudes, one can separate all the various effects, again
obtaining clean CKM phase information. For example, in the case of
$B^0\to\pi^+\pi^-$, it is necessary to measure the rates for all possible
charge states in $B\to\pi\pi$ and ${\bar B}\to\pi\pi$, as well as the time
dependence of $B^0(t)\to\pi^+\pi^-$ \cite{isospin}. A similar analysis can
be done for the decays $B\to\pi K$ \cite{piKisospin}.

Given that isospin symmetry is such a useful tool in the measurement of CP
violation in the $B$ system, it is only natural to next ask whether
considerations of flavor SU(3) symmetry \cite{Zeppenfeld}-\cite{GHLR} can
lead to anything interesting. It is the purpose of this Letter to show that
indeed they do: by using SU(3) and making some reasonable approximations it
is possible to obtain not only the weak phases $\alpha$ and $\gamma$, but
also strong phase information and the sizes of different diagrams. This is
done through the measurement of several $B$ decay rates to $\pi\pi$, $\pi
K$ and $K {\bar K}$, most of whose branching ratios are of the order of
$10^{-5}$. {\it No time-dependent measurements are needed.} (If
time-dependence can be measured, then in addition the weak angle $\alpha$
can be obtained in other, independent ways.) There are several independent
ways to obtain all this information. The use of these gives a possible way
to reduce the discrete ambiguities and also tests the SU(3) flavor
symmetry.

\bigskip
\centerline{\bf II. SU(3) RELATIONS AMONG AMPLITUDES}
\bigskip

We apply flavor SU(3) symmetry to the decays of $B$ mesons to pairs of
light pseudoscalar mesons $\pi\pi$, $\pi K$ and $K {\bar K}$. The SU(3)
amplitudes for these decays, which involve all possible charge states, can
be expressed in terms of the following diagrams (see Fig.~1): a ``tree''
amplitude $T$ or $T'$, a ``color-suppressed'' amplitude $C$ or $C'$, a
``penguin'' amplitude $P$ or $P'$, an ``exchange'' amplitude $E$ or $E'$,
an ``annihilation'' amplitude $A$ or $A'$, and a ``penguin annihilation''
amplitude $PA$ or $PA'$. Here an unprimed amplitude stands for a
strangeness-preserving decay, while a primed contribution stands for a
strangeness-changing decay. As noted in Refs.~\cite{Zeppenfeld,GRL,GHLR},
this set of amplitudes is over-complete. The physical processes of interest
involve only five distinct linear combinations of these six terms.

The diagrams denoted by $E$, $A$ and $PA$ involve contributions to
amplitudes which should behave as $(f_B/m_B)$ in comparison with those from
the diagrams $T$, $C$ and $P$ (and similarly for their primed
counterparts). This suppression is due to the smallness of the $B$ meson
wavefunction at the origin and should remain valid unless rescattering
effects are important. Such rescatterings indeed could be responsible for
certain decays of charmed particles, but should be less important for the
higher-energy $B$ decays. In addition the diagrams $E$ and $A$ are also
helicity suppressed by $(m_{u,d,s}/m_B)$ since the $B$ mesons are
pseudoscalars. Neglecting the contributions of the above diagrams, we are
left with the 6 diagrams $T$, $T'$, $C$, $C'$, $P$ and $P'$. These six
complex parameters determine the 13 allowed $B$ decays to states with pions
and kaons, as listed in Table 1. Following the conventions in
Refs.~\cite{Zeppenfeld,GRL,GHLR}, we take the $u$, $d$, and $s$ quark to
transform as a triplet of flavor SU(3), and the $-\bar u$, $\bar d$, and
$\bar s$ to transform as an antitriplet. Thus the $\pi$-mesons and kaons
form part of an octet and are defined as $\pi^+ \equiv u \bar d$, $\pi^0
\equiv (d \bar d - u \bar u)/\s$, $\pi^- \equiv - d \bar u$, $K^+ \equiv u
\bar s$, $K^0 \equiv d \bar s$, $\bar K^0 \equiv s \bar d$ and $K^- \equiv
- s \bar u$. The $B$ mesons, which are in the triplet or anti-triplet
representation, are taken to be $B^+ \equiv \bar b u$, $B^0 \equiv \bar b
d$, $B_s \equiv \bar b s$, $B^- \equiv - b\bar u$, $\Bbar^0 \equiv b \bar
d$ and $\Bbar_s \equiv b\bar s$.

\begin{table}
\caption{ \er
The 13 decay amplitudes in terms of the 8 graphical combinations. The $\sst
\protect\sqrt{2}(B^+\to\pi^+\pi^0)$ in the $\sst -(T+C)$ column means that
$\sst A(B^+\to\pi^+\pi^0) = -(T+C)/\protect\sqrt{2}$, and similarly for
other entries. Processes in the same column can be related by an amplitude
equality, e.g.~the amplitudes for $\sst B^+\to K^+\Kbar^0$ and $\sst B^0\to
K^0\Kbar^0$ are equal.}
\begin{center}
\begin{tabular}{l l c c c c c c} \hline
$~~~-(T+C)$ & $~~~-(C-P)$         & $-(T+P)$     &
  $(P)$               & \\ \hline
$\s(B^+\to\pi^+\pi^0)$  & $\s(B^0\to\pi^0\pi^0)$ & $B^0\to\pi^+\pi^-$ &
$B^+\to K^+ \Kbar^0$ & \\
                    & $\s(B_s\to\pi^0 \Kbar^0)$ & $B_s\to\pi^+K^-$   &
$B^0\to K^0 \Kbar^0$  &  \\ \hline
$~-(T'+C'+P')$ & $~~~-(C'-P')$ & $-(T'+P')$     &
  $(P')$               & \\ \hline
$\s(B^+\to\pi^0 K^+)$  & $\s(B^0\to\pi^0 K^0)$ & $B^0\to\pi^- K^+$ &
$B^+\to\pi^+ K^0$ & \\
                       &                   & $B_s\to K^- K^+$  &
$B_s\to K^0 \Kbar^0$  &  \\ \hline
\end{tabular}
\end{center}
\end{table}

The primed and unprimed diagrams are not independent, but are related by
CKM matrix elements. In particular, $T'/T = C'/C = r_u$, where $r_u \equiv
\v u s/ \v u d\approx 0.23$. Assuming that the penguin amplitudes are
dominated by the top quark loop \cite{toppenguin}, one has $P'/P = r_t$,
with $r_t \equiv \v t s/ \v t d$. We therefore have 13 decays described by
3 independent graphs, implying that there are 10 relations among the
amplitudes. These can be expressed in terms of 6 amplitude equalities, 3
triangle relations, and one quadrangle relation. We will soon see how to
replace the quadrangle relation by a triangle one.

The three independent triangle relations and one quadrangle relation are
\beq
\label{trita}
\s A(B^+ \to \pi^+ \pi^0) = \s A(B^0 \to \pi^0 \pi^0) +
A(B^0 \to \pi^+ \pi^-) ~~~,
\eeq
\beq
\label{tritaa}
\s A(B^+ \to \pi^+ \pi^0) = {1\over r_u} \s A(B^0 \to \pi^0 K^0) +
{1\over r_u} A(B^0 \to \pi^- K^+) ~~~,
\eeq
\beq
\label{tritb}
\s A(B^+ \to \pi^+ \pi^0) = {1\over r_u} \s A(B^+ \to \pi^0 K^+) +
{1\over r_u} A(B^+ \to \pi^+ K^0) ~~~,
\eeq
\beq
\label{tritc}
A(B^0 \to \pi^- K^+) + A(B^+ \to \pi^+ K^0) =
r_u [ A(B^0 \to \pi^+ \pi^-) + A(B^+ \to K^+ \bar K^0) ].
\eeq
We have chosen to use only decays of $B^0$ or $B^+$ mesons in these
relations -- other decays (of $B_s$, for example) can be substituted using
amplitude equalities. Schematically, these four relations can be written as
\beq
\label{xyz}
(T+C)=(C-P)+(T+P)
\eeq
\beq
\label{xypzp}
(T+C)=(C'-P')/r_u + (T'+P')/r_u
\eeq
\beq
\label{xwp}
(T+C) = (T'+C'+P')/r_u - (P')/r_u
\eeq
\beq
\label{zp}
(T'+P') - (P') = r_u (T+P) - r_u(P)
\eeq

\bigskip
\centerline{\bf III. MEASURING WEAK AND STRONG PHASES }
\bigskip

The amplitude for $B\to\pi^+\pi^0$ decay, given by $-(T+C)/\s$, is pure
$\Delta I = 3/2$. Hence $(T+C)$ has only one term, which we denote by
$A_{\sss I=2} e^{i\phi_2} e^{i\delta_2}$, and we write the triangle
relations as:
\beq
A_{\sss I=2} e^{i\phi_2} e^{i\delta_2}
=(A_C e^{i\phi_{C}}e^{i\delta_C}-A_P e^{i\phi_P}e^{i\delta_P})
+(A_T e^{i\phi_T}e^{i\delta_T}+A_P e^{i\phi_P}e^{i\delta_P})~~~,
\eeq
\beq
A_{\sss I=2} e^{i\phi_2} e^{i\delta_2}
=(A_{C'} e^{i\phi_{C'}}e^{i\delta_{C'}}
-A_{P'} e^{i\phi_{P'}} e^{i\delta_{P'}})/r_u
+(A_{T'} e^{i\phi_{T'}}e^{i\delta_{T'}}
+A_{P'} e^{i\phi_{P'}} e^{i\delta_{P'}})/r_u~~,
\eeq
\beq
A_{\sss I=2} e^{i\phi_2} e^{i\delta_2}
=(A_{T'} e^{i\phi_{T'}}e^{i\delta_{T'}} + A_{C'}
e^{i\phi_{C'}}e^{i\delta_{C'}}
+A_{P'} e^{i\phi_{P'}} e^{i\delta_{P'}})/r_u
-A_{P'} e^{i\phi_{P'}} e^{i\delta_{P'}}/r_u~~~,
\eeq
where the $\phi_i$ are the weak phases and the $\delta_i$ are the strong
phases. The $\delta_i$ are chosen such that the quantities $A_{\sss I=2}$,
$A_T$, $A_{T'}$, $A_C$, $A_{C'}$, $A_P$ and $A_{P'}$ are real and positive.
By SU(3) symmetry the strong phases for the primed graphs are the same as
those for the unprimed ones. Working with the Wolfenstein parametrization
\cite{Wolfenstein} of the CKM matrix, it is easy to see that the weak
phases of the various amplitudes are: $\phi_2= \phi_T = \phi_{T'} = \phi_C
= \phi_{C'} = \gamma$, $\phi_P=-\beta$, and $\phi_{P'} = \pi$ (up to
corrections of order $\lambda^2 \approx 0.05$). Also, $A_{T'}/r_u=A_T$ and
$A_{C'}/r_u=A_C$. Finally, multiplying through on both sides by
$\exp(-i\gamma-i\delta_2)$, the 3 triangle relations become
\begin{eqnarray}
\label{taa}
A_{\sss I=2} & = &
(A_C e^{i\Delta_C} + A_P e^{i\alpha} e^{i\Delta_P})
+(A_T e^{i\Delta_T} - A_P e^{i\alpha} e^{i\Delta_P}), \\
\label{tb}
A_{\sss I=2} & = &
(A_C e^{i\Delta_C} + A_{P'} e^{-i\gamma}e^{i\Delta_P}/r_u)
+(A_T e^{i\Delta_T} - A_{P'} e^{-i\gamma}e^{i\Delta_P}/r_u), \\
\label{tc}
A_{\sss I=2} & = & (A_T e^{i\Delta_T} + A_C e^{i\Delta_C}
- A_{P'} e^{-i\gamma}e^{i\Delta_P}/r_u)
+ A_{P'} e^{-i\gamma}e^{i\Delta_P}/r_u),
\end{eqnarray}
where we have defined $\Delta_i \equiv \delta_i - \delta_2$. {\it These
three triangles and their charge conjugates can be used to determine all
weak phases, all strong phase differences, and the sizes of the various
diagrams.}

Consider first the two triangle relations in Eqs.~(\ref{tb}) and
(\ref{tc}). These relations define two triangles which share a common base.
Each triangle is determined up to a two-fold ambiguity, since it can be
reflected about its base. Implicit in these two triangle relations is the
relation
\beq
\label{cttri}
A_{\sss I=2} = |T + C| = A_T e^{i\Delta_T} + A_C e^{i\Delta_C}~~~.
\eeq
Thus both of these triangles also share a common subtriangle with sides
$T+C$, $C$ and $T$ as shown in Fig.~2(a). The key point is this: the
subtriangle is completely determined, up to a four-fold ambiguity, by the
two triangles in Eqs.~(\ref{tb}) and (\ref{tc}). This is because both the
magnitude and relative direction of $P'/r_u$ are completely determined by
constructing the triangle in Eq.~(\ref{tc}). Therefore the point where the
vectors $C$ and $T$ meet is given by drawing the  vector $P'/r_u$ from the
vertex opposite the base [see Fig.~2(a)]. (A similar construction would
have given the same point if we had used the vector $T+P'/r_u$ instead of
$P'/r_u$.) Thus, if we measure the five rates for

\vspace*{2mm}
$B^0 \to \pi^0 K^0$ (giving $|C - P'/r_u|$),

\vspace*{2mm}
$B^0 \to \pi^-K^+$ (giving $|T + P'/r_u|$),

\vspace*{2mm}
$B^+ \to \pi^0 K^+$ (giving $|T + C + P'/r_u|$),

\vspace*{2mm}
$B^+ \to \pi^+ K^0$ (giving $|P'/r_u|$), and

\vspace*{2mm}
$B^+ \to \pi^+\pi^0$ (giving $|T + C|=A_{\sss I=2}$,
i.e.~the triangle's base),

\vspace*{2mm}
\noindent
we can determine $\Delta_P - \gamma$, $|T|$ and $|C|$, up to a two-fold
ambiguity and $\Delta_C$ and $\Delta_T$ up to a four-fold ambiguity. As we
will discuss later, these discrete ambiguities can be at least partially
removed through the knowledge of the relative magnitudes of $|P|$, $|C|$,
$|T|$ and $|P'|$, and through independent measurements of the amplitudes
and the strong and weak phases.

If we also measure the rates for the CP-conjugate processes of the above
decays, we can get more information. These CP-conjugate decays obey similar
triangle relations to those in Eqs.~(\ref{tb}) and (\ref{tc}). However,
recall that under CP conjugation, the weak phases change sign, but strong
phases do not. Thus we can perform an identical analysis with the
CP-conjugate processes, giving us another, independent determination of
$|T|$, $|C|$, $\Delta_C$ and $\Delta_T$. But, instead of $\Delta_P -
\gamma$, this time we get $\Delta_P + \gamma$. Thus we obtain $\Delta_P$
and $\gamma$ separately. Note that it is not, in fact, necessary to measure
all 5 CP-conjugate processes. The rate for $B^- \to \pi^-\pi^0$ is the same
as that for $B^+ \to \pi^+\pi^0$, since they involve a single weak phase
and a single strong phase. Similarly, the rates for $B^+ \to \pi^+ K^0$ and
$B^- \to \pi^- K^0$ are equal. Therefore, in order to extract $\gamma$, in
addition to the above 5 rates, we need only measure

\vspace*{2mm}
$\Bbar^0 \to \pi^0 \Kbar^0$ (giving $|\bar C - \bar{P'}/r_u|$),

\vspace*{2mm}
$\Bbar^0 \to \pi^+K^-$ (giving $|\bar T + \bar{P'}/r_u|$), and

\vspace*{2mm}
$B^- \to \pi^0 K^-$ (giving $|\bar T + \bar C + \bar{P'}/r_u|$).

\vspace*{2mm}
\noindent
To sum up, by measuring the above 8 rates, the following quantities can be
obtained: the weak phase $\gamma$, the strong phase differences $\Delta_T$,
$\Delta_C$ and $\Delta_P$, and the magnitudes of the different amplitudes
$|T|,~|C|$ and $|P'|$.

A few comments are worth making regarding the above two-triangle
construction. First, note that all measurements are time-independent, and
that no observation of CP violation is required to obtain the various
quantities. This construction extends that of Ref.~\cite{GRL}, in which it
is observed that the measurement of the sides of the triangle in
Eq.~(\ref{tc}) and its CP-conjugate can be used to obtain the weak angle
$\gamma$. The second point is that most of the decays are self-tagging, the
only exceptions being $B^0 \to \pi^0 K^0$ and $\Bbar^0 \to \pi^0 \Kbar^0$.
However, even at a symmetric $e^+e^-$ machine operating at the
$\Upsilon(4S)$, where it is not possible to tag individual $B$'s, these two
rates can still be obtained. By measuring the two time-integrated rates for
the $\bd\bdb$ pair to decay to the final state $\pi^0 K_S$ plus a
semileptonic tag $[(D\ell{\bar\nu}X)_{tag}$ or $({\bar D} {\bar\ell} \nu
{\bar X})_{tag}]$, the two rates can be extracted. Finally, if
time-dependent measurements are possible, one can independently measure
$\alpha$ through CP violation in neutral $B$ decays to $\pi^0 K_S$
\cite{piKisospin}.

Now consider the triangle relations in Eqs.~(\ref{taa}) and (\ref{tb}).
These two triangles share a common base with each other and also with the
sub-triangle in Eq.~(\ref{cttri}) (which still holds). Unlike the previous
two-triangle construction, however, the shape of the sub-triangle is not
yet fixed. Nevertheless, the point where the vectors $C$ and $T$ meet can
still be determined [see Fig.~2(b)]. This point is connected to the apex of
the triangle in Eq.~(\ref{taa}) by the vector $P$, and to the apex of the
triangle in  Eq.~(\ref{tb}) by the vector $P'$. Since the magnitudes of the
penguin diagrams $P$ and $P'$ are measured by the rates of $B^+\to
K^+\Kbar^0$ and $B^+\to\pi^+K^0$, respectively, the meeting point of $C$
and $T$ is determined by the intersection of the two circles of Fig.~2(b).
Thus, the sub-triangle is completely determined up to an eight-fold
ambiguity. This eight-fold ambiguity correspond to the two possible
intersections of the circles, in addition to the 2 two-fold ambiguities
caused by reflecting each triangle about its base. Thus by measuring the 7
rates

\vspace*{2mm}
$B^+\to K^+\Kbar^0$ (giving $|P|$),

\vspace*{2mm}
$B^+\to\pi^+K^0$ (giving $|P'|$),

\vspace*{2mm}
$B^0\to\pi^0\pi^0$ (giving $|C-P|$),

\vspace*{2mm}
$B^0\to\pi^+\pi^-$ (giving $|T+P|$),

\vspace*{2mm}
$B^0\to\pi^0 K^0$ (giving $|C-P'/r_u|$),

\vspace*{2mm}
$B^0\to \pi^-K^+$ (giving $|T+P'/r_u|$), and

\vspace*{2mm}
$B^+\to\pi^+\pi^0$ (giving $|T + C|$),

\vspace*{2mm}
\noindent
we can extract $\Delta_P+\alpha$, $\Delta_P - \gamma$, $\Delta_C,$ and
$\Delta_T$, up to an eight-fold ambiguity, and $|T|$ and $|C|$ up to a
four-fold ambiguity. Through the two quantities  $\Delta_P+\alpha$ and
$\Delta_P - \gamma$, we can then determine the weak phase $\beta$ (using
$\beta=\pi-\alpha-\gamma$), up to discrete ambiguities. As in the first
two-triangle construction, all rates are time-independent. What is
surprising, perhaps, about this particular construction is that {\it it is
not even necessary to measure the CP-conjugate rates in order to obtain
$\beta$.} The reason is that SU(3) flavor symmetry implies the equality of
the strong final-state phases of two different amplitudes, in this case $P$
and $P'$. Subtracting the (strong plus weak) phase of one amplitude from
the other then determines a weak phase. Usually, in a given process,
without measuring the charge-conjugate rate one can only measure the sum of
a weak and a strong phase.

If the CP-conjugate rates are also measured, we can obtain $\Delta_P$,
$\alpha$, and $\gamma$ separately. This provides another, independent
determination of $|T|$, $|C|$, $\Delta_C$ and $\Delta_T$. As in the first
construction, no observation of CP violation is necessary to make such
measurements. Again, it is not necessary to measure all the CP-conjugate
rates -- only the following four can be different from their counterparts:

\vspace*{2mm}
$\Bbar^0\to\pi^0\pi^0$ (giving $|\bar C-\bar P|$),

\vspace*{2mm}
$\Bbar^0\to\pi^+\pi^-$ (giving $|\bar T+\bar P|$),

\vspace*{2mm}
$\Bbar^0\to\pi^0 \Kbar^0$ (giving $|\bar C-\bar{P'}/r_u|$), and

\vspace*{2mm}
$\Bbar^0\to \pi^+K^-$ (giving $|\bar T+\bar{P'}/r_u|$).

\vspace*{2mm}
\noindent
If time-dependent measurements can be performed, then mixing-induced CP
violation can be seen in neutral $B$ decays to $\pi^+\pi^-$ or $\pi^0 K_S$.
The measurement of such CP violation, combined with the measurements of the
penguin diagram, will give another, independent value for the angle
$\alpha$, just as in the isospin analysis \cite{isospin,piKisospin}.

Finally, the third two-triangle construction uses the triangle relations in
Eqs.\ (\ref{taa}) and (\ref{tc}). This construction is almost identical to
the previous one. The only difference is that the decay $B^0\to\pi^0 K^0$
(giving $|C-P'/r_u|$) is replaced by the decay $B^+\to\pi^0 K^+$ (giving
$|T+C+P'/r_u|$). (This is experimentally preferable since $B^+$ decays are
self-tagging.) Therefore, while the radius of one of the circles is still
$|P|$ as in Fig.~2(b), the radius of the other circle is given by
$|T+P'/r_u|$. Although this construction does not provide new information,
it can nevertheless be used as an independent measurement of the weak
phases, the strong phase differences, and the size of the various diagrams.

An interesting feature of the last two constructions is that the
quadrilateral symbolizing the quadrangle relation of Eq.~(\ref{zp}) is
contained in the figure. This might seem to imply that it is not an
independent relation. In fact, this is not so. What has happened is that
the amplitude relation of Eq.~(\ref{zp}) has been implicitly replaced by
the triangle relation Eq.~(\ref{cttri}), which is a relation between an
amplitude and two graphs.

The three constructions use $B$ decays to $\pi\pi$, $\pi K$ and $K{\bar K}$
final states. At present, the decays $B^0 \to \pi^+\pi^-$ and/or $\pi^-
K^+$ have been observed, but the two final states cannot be distinguished
\cite{CLEO}. The combined branching ratio is about $2\times 10^{-5}$.
Assuming equal rates for $\pi^+\pi^-$ and $\pi^- K^+$, which seems likely,
the amplitudes $|T|$ and $|P'|$ should be about the same size. On the other
hand, the amplitude $|C|$ is expected to be about a factor of 5 smaller:
the amplitudes $|T|$ and $|C|$ are basically the same as $|a_1|$ and
$|a_2|$, respectively, introduced in Ref.~\cite{BSW}, for which the values
$|a_1| = 1.11$ and $|a_2| = 0.21$ have been found \cite{Lindner}. The ratio
$|P/T|$ has also been estimated to be small, $\lsim 0.20$
\cite{gropenguin}. Therefore all the decays used in these constructions
should have branching ratios of the order of $10^{-5}$, with the exception
of $B\to K{\bar K}$ ($P$) and $B^0\to\pi^0\pi^0$ [$\sim (C-P)$], which are
probably an order of magnitude smaller.

The knowledge that the amplitudes obey the hierarchy
$|P|,~|C|<|T|<|P'/r_u|$ will also help in reducing discrete ambiguities.
For example, in the first two-triangle construction [Fig.~2(a)], we noted
in the discussion following Eq.~(\ref{cttri}) that the subtriangle can be
determined up to a four-fold ambiguity. However, two of these four
solutions imply that $|C|$ and $|T|$ are both of order $|P'/r_u|$, which
violates the above hierarchy. Thus the four-fold ambiguity in the
determination of the subtriangle is reduced to a two-fold ambiguity, and
the discrete ambiguities in the determination of subsequent quantities such
as $\Delta_P-\gamma$, $\Delta_C$, etc., are likewise reduced. The
ambiguities in the other two constructions can be partially removed in a
similar way.

All three two-triangle constructions described above rely on two
assumptions. The first is that the diagrams $A$, $E$ and $PA$ (and their
primed counterparts) can be neglected. This can be tested experimentally.
The decays $B^0 \to K^+ K^-$ and $B_s \to \pi^+ \pi^-$ can occur only
through the diagrams $E$ and $PA$, and $E'$ and $PA'$, respectively.
Therefore, if the above assumption is correct, the rates for these two
decays should be much smaller than the rates for the decays in Table 1.

The second assumption is that of an unbroken SU(3) symmetry. We know,
however, that SU(3) is in fact broken in nature. Assuming factorization,
SU(3)-breaking effects can be taken into account by including the meson
decay constants $f_\pi$ and $f_K$ in the relations between $B\to\pi\pi$
decays and $B\to\pi K$ decays \cite{Silvawolf}. In other words, the factor
$r_u$ which appears in two of the triangle relations should be multiplied
by $f_K/f_\pi\approx 1.2$. One way to test whether this properly accounts
for all SU(3)-breaking effects is through the rate equalities in Table 1.
Even if it turns out that $f_K/f_\pi$ does not take into account all
SU(3)-breaking effects, the large number of independent measurements is
likely to help in reducing uncertainties due to SU(3) breaking. For
example, note that, not counting the CP-conjugate processes, the last two
constructions have six of their seven rates in common. This means that a
measurement of only eight decay rates gives two independent measurements of
$|T|$, $|C|$, $\Delta_C$, $\Delta_T$, $\Delta_P - \gamma$ and $\Delta_P +
\alpha$. In fact, these eight rates already contain the five rates of the
first construction [Fig.~2(a)]. Thus we actually have three independent
ways of arriving at $|T|$, $|C|$, $\Delta_C$, $\Delta_T$ and $\Delta_P -
\gamma$. Including also the CP-conjugate processes, we have a total of 13
$B$-decay rate measurements which give us six independent ways to measure
$|T|$, $|C|$, $\Delta_C$ and $\Delta_T$, five ways to measure $\Delta_P$,
three independent ways to measure $\gamma$, and two ways to measure
$\alpha$. (If time-dependent measurements are possible, there are
additional independent ways to measure $\alpha$.) The point is that the
three two-triangle constructions include many ways to measure the same
quantity. This redundancy provides a powerful way to test the validity of
our SU(3) analysis and reduces the discrete ambiguities in the
determination of the various quantities.

\bigskip
\centerline{\bf IV. CONCLUSIONS}
\bigskip

We have presented an analysis based on three triangle relations involving
$B$ decay amplitudes to $\pi\pi$, $\pi K$ and $K {\bar K}$ which are a
consequence of the SU(3) flavor symmetry of the strong interactions and the
smallness of annihilation-like diagrams. These relations permit the
extraction of the weak phases $\alpha$ and $\gamma$ even if penguin
contributions are substantial. In addition, one obtains the strong phase
differences and the size of the individual diagrams. No time-dependent
measurements are needed, nor is it necessary to observe CP violation to
determine these quantities. If time-dependent measurements are possible,
additional, independent measurements of $\alpha$ can also be made. Most
branching ratios are expected to be of the order of $10^{-5}$, although a
few could be an order of magnitude smaller. Interestingly, in some cases
our method provides a measurement of CP-violating weak phases without
necessarily measuring CP-conjugate processes. All quantities are measured
up to certain discrete ambiguities. However, the method includes many
independent measurements of the same quantities, which can be used to
considerably reduce the discrete ambiguities. Furthermore, this redundancy
is likely to be of great help in evaluating the size of SU(3)-breaking
effects.
\newpage

\centerline{\bf ACKNOWLEDGMENTS}
\bigskip
We thank L. Wolfenstein for helpful discussions. M. Gronau and J. Rosner
respectively wish to acknowledge the hospitality of the Universit\'e de
Montr\'eal and the Technion during parts of this investigation. This work was
supported in part by the United States -- Israel Binational Science Foundation
under Research Grant Agreement 90-00483/2, by the German-Israeli Foundation for
Scientific Research and Development, by the Fund for Promotion of Research at
the Technion, by the N.S.E.R.C. of Canada and les Fonds F.C.A.R. du Qu\'ebec,
and by the United States Department of Energy under Contract No. DE FG02
90ER40560.

\newpage

\newpage

\centerline{\bf FIGURE CAPTIONS}
\bigskip

\noindent
FIG.\ 1. Diagrams describing decays of $B$ mesons to pairs of light
pseudoscalar mesons. Here $\bar q = \bar d$ for unprimed amplitudes and
$\bar s$ for primed amplitudes. (a) ``Tree'' (color-favored) amplitude $T$
or $T'$; (b) ``Color-suppressed'' amplitude $C$ or $C'$; (c) ``Penguin''
amplitude $P$ or $P'$ (we do not show intermediate quarks and gluons); (d)
``Exchange'' amplitude $E$ or $E'$; (e) ``Annihilation'' amplitude $A$ or
$A'$; (f) ``Penguin annihilation'' amplitude $PA$ or $PA'$.
\bigskip

\noindent
Fig.\ 2. Triangle relations used to obtain weak phases and strong
final-state phase shift differences. The black dot corresponds to the
solution for the vertex of the triangle in Eq.~(\ref{cttri}). (a) Relation
based on Eqs.~(\ref{tb}) (upper triangle) and (\ref{tc}) (lower triangle).
(b) Relation based on Eqs.~(\ref{taa}) (lower triangle with small circle
about its vertex) and (\ref{tb}) (upper triangle with large circle about
its vertex). The relation based on (\ref{taa}) and (\ref{tc}) follows an
almost identical construction. One possible set of decay processes which
can be used to construct these triangles is given in Eqs.~(\ref{trita}),
(\ref{tritaa}) and (\ref{tritb}).

\end{document}